\documentclass[10pt]{llncs}
\usepackage[title]{appendix}

\usepackage{cite}
\usepackage{appendix}
\usepackage{amsmath,amssymb,amsfonts}

\usepackage{algorithm,algpseudocode}
\usepackage{graphicx}
\usepackage{textcomp}
\usepackage{bmpsize}
\usepackage{xcolor}
\usepackage{lipsum}
\usepackage[colorlinks=true,urlcolor=black]{hyperref}

\usepackage{multirow}
\usepackage{subcaption}
\usepackage{graphicx}
\usepackage{lipsum}
\usepackage{tcolorbox}
\usepackage{framed}
\usepackage{svg}
\usepackage{outlines}

\usepackage{verbatim}

\newcommand{\scbf}[1]{\vspace {0.05in}\noindent{\textbf{#1}}}

\begin{document}

\title{Empirical Analysis of Evasion and Poisoning Against Malware Data Drift Detection}

\author{Mingyue Yang\orcidID{0009-0002-7783-1091} \and David Lie\orcidID{0000-0002-2000-6827} \and Nicolas Papernot\orcidID{0000-0001-5078-7233}}
\institute{University of Toronto and Vector Institute\\
\email{myshirley.yang95@gmail.com,\{david.lie,nicolas.papernot\}@utoronto.ca}}
\maketitle

\begin{abstract}
As concept drift due to malware evolution presents challenges for malware classification, machine learning-based data drift detection tools are developed to mitigate this problem. These data drift detector tools are designed for a different purpose and built with different techniques compared to malware classifiers. Although evasion and poisoning attacks against machine learning-based malware classifiers can cause misclassification of malware samples, it is not clear how these attacks work against data drift detectors and malware classifiers in combination. This work investigates the effect of evasion and poisoning attacks on the data drift detector along with the malware classifier. We demonstrate how unique characteristics of data drift detectors cause attacks against malware classifiers to work differently against them.
\end{abstract}

\section{Introduction}
Machine learning models can detect whether a piece of software is malware or not~\cite{raff2018malware, arp2014drebin, anderson2018ember}. These machine learning malware classifiers have better generalizability than signature-based malware detection approaches and can catch previously unseen patterns. However, as both malware and benign software (benignware) evolve over time, machine learning-based malware classifiers can soon become outdated and will no longer perform well on new inputs. To mitigate this problem, machine learning-based data drift detectors~\cite{yang2021cade, jordaney2017transcend, barbero2022transcending} have been proposed to find drifting samples for malware classification (i.e. samples that drift away from the data distribution of the training set). These data drift detectors can be used to reject the predictions of samples that are likely misclassified~\cite{barbero2022transcending} or to continuously train the malware classifier using detected drifting samples to overcome concept drift~\cite{chen2023continuous}. As these malware data drift detectors have different goals than malware classification, they have different loss terms and/or model architectures than the malware classifiers~\cite{yang2021cade, jordaney2017transcend, barbero2022transcending}. These malware data drift detectors behave differently from the malware classifier models they work with.

As machine learning models are susceptible to adversarial attacks, many existing works study evasion and poisoning attacks against machine learning malware classifiers~\cite{boutsikas2021evading, lucas2021malware, demetrio2021adversarial, yang2023jigsaw, severi2021explanation}. However, a successful attack for such system should work against \textbf{\textit{both}} the malware classifier \textbf{\textit{and}} the \textbf{\textit{data drift detector}}.

As it is unclear how malware data drift detectors, also built with machine learning models but have different designs and purposes, behave under evasion and poisoning attacks, our paper makes the following contributions:

1) We conduct evasion and poisoning attacks against the data drift detector along with the malware classifier. We study how factors such as the size of the perturbations and the number of poisoning samples affect data drift detectors with different underlying model architectures.

2) We demonstrate properties of data drift detectors that cause the attacks to work differently against them. We illustrate cases for which attacks that work for the malware classifier do not work well for data drift detectors.

\section{\label{sec:background_data_drift}Background: Data Drift Detector}

Our work evaluates attacks against two recent data drift detectors designed for malware as follows.

\scbf{CADE~\cite{yang2021cade}}: The CADE data drift detector~\cite{yang2021cade} trains a contrastive autoencoder (CAE) model that converts data from the input space to a lower dimensional embedding space. The loss function of the CAE consists of a \textbf{\textit{contrastive loss}} term that pushes samples from different classes away from each other and keeps samples from the same class closer together in the embedding space. This separates the benign samples and malware samples in the embedding space. Similar to other autoencoders, the CAE model also has a \textbf{\textit{reconstruction loss}} term that minimizes the difference between the reconstructed input from its decoder and the raw input to its encoder. The reconstruction loss ensures the CAE embedding space accurately represents details in the input. Euclidean distance in the CAE embedding space denotes how similar samples are, and samples closer to each other are more similar.

To determine whether a tested sample is a drifting sample of a given class $i$, CADE first finds the centroid of that class in its CAE embedding space ($centroid_{i}$), which is the average of embedding space representation of all training samples in that class. CADE computes the distance in embedding space between this tested sample and the class centroid: $d_{i}^{test} = |CAE\_latent(sample^{test}) - centroid_{i}|$, and considers samples closer to the class centroid as a less drifted sample of that class.

CADE also finds the median distance to the class centroid for all training samples in the given class $i$: $\tilde{d}_{i}$. With these results, CADE then calculates an anomaly score that compares the distance for the tested sample and the median distance for all training samples in the class $i$: $anomaly_{i}^{test} = \frac{|d_{i}^{test} - \tilde{d}_{i}|}{b*\tilde{d}_{i}}$, where $b$ is a fixed constant from CADE~\cite{yang2021cade}. This distance-based \textbf{\textit{anomaly score}} can be used to find drifting samples of a specific class. For a data sample, a higher anomaly score for a specific class means this sample is more likely to be a drifting sample for that class. The \textbf{\textit{OOD (out-of-distribution) score}} of a sample is the minimum of its anomaly scores for all classes. A sample with a high OOD score is a drifting sample for all classes.

\scbf{Transcend and Transcendent~\cite{barbero2022transcending}}: Transcend~\cite{jordaney2017transcend} uses p-values from conformal prediction to determine whether a sample is a drifting sample of a class $i$. The p-value of a sample $s$ for a given class $i$ is calculated as $p_{s}^{I} = \frac{ |\{ s_{i} \in I | sim(s_{i}, I \setminus s_{i}) \le sim(s, I) \}| }{|I|}$, where $I$ is the set of all samples in class $i$. It is the proportion of training samples in class $i$ that are at least as dissimilar to other samples in class $i$ as the given sample, with the similarity measured using a similarity metric $sim$.

While different similarity metrics can be used to compute p-values, we use the similarity metrics in Transcend that can work with either an MLP model (Multi-Layer Perceptron) or a CAE model~\cite{chen2023continuous, barbero2022transcending}. To compute p-values, a linear SVM model is first built upon either the embedding space of the CAE model or the last latent layer of the MLP right before the output layer of the MLP model, which best represents the decision-making of the MLP right before its prediction. With the built SVM, the distance to the hyperplane of the SVM decision boundary is used as similarity metrics to compute p-values: a larger distance to the hyperplane means the sample is more similar to other samples in the given class $i$. For samples not predicted as the given class $i$ by the SVM (i.e. these samples are on the other side of the SVM hyperplane), their distances to the hyperplane are negated, so these samples have negative distance measures.

Transcend~\cite{jordaney2017transcend} then computes a \textbf{\textit{credibility score}}, a \textbf{\textit{confidence score}}, and a \textbf{\textit{cred+conf score}} for using the p-value of a given sample. For a predicted class $i$, the \textbf{\textit{credibility score}} of a sample is simply its p-value for class $i$. A high credibility score means this sample is more similar to other samples in class $i$. The \textit{\textbf{confidence score}} for class $i$, however, is calculated as 1 minus the highest p-value for classes other than the class $i$. A high confidence score means this sample is not similar to classes other than class $i$. A non-drifting sample of class $i$ would have both a high credibility score and a confidence score, meaning it is only similar to class $i$ while not similar to other classes. To ensure good confidence and credibility, a \textit{\textbf{cred+conf score}}, calculated by multiplying the credibility score by the confidence score, represents the driftedness of a sample. Samples with lower cred+conf scores are more drifted for the given class $i$.

Our work uses Transcendent~\cite{barbero2022transcending}, an improved version of Transcend~\cite{jordaney2017transcend}, which approximates and sometimes surpasses the performance of the initial Transcend work while requiring less computational overhead.

\section{\label{sec:general_threat_model_universal_attack}Threat Model}

\scbf{Datasets}: We use the Androzoo dataset~\cite{allix2016androzoo} and the Bodmas dataset~\cite{yang2021bodmas} with malware and benignware spanning over several months. In the Androzoo dataset, all features must be either 0 or 1, indicating whether a component is present in the sample. In the Bodmas dataset, different features have different ranges of numerical values. The features of both datasets can be obtained from simple static analysis without running the malware samples.

\scbf{Victim Models and Drifting Sample Selection}: We assume the victim starts with initial malware classifier and data drift detector models trained using the \textbf{\textit{initial training set}} in Table~\ref{tab:bodmas_androzoo_dataset}. These initial models are trained with data balancing by randomly oversampling malware samples. The victim uses MLP for malware classification, and either CADE~\cite{yang2021cade} or Transcendent~\cite{barbero2022transcending} for data drift detection (See Section~\ref{sec:background_data_drift}). As new samples from the \textbf{\textit{future selection set}} (See Table~\ref{tab:bodmas_androzoo_dataset}) are encountered, the malware classifier predicts whether these encountered samples are malware or not.

\begin{table*}
\vspace{-30pt}
\caption{\label{tab:bodmas_androzoo_dataset} Bodmas and Androzoo Datasets, Without Data Balancing}
\resizebox{\textwidth}{!}{
\begin{tabular}{|c|c|c|c|c|}
\hline
& \multicolumn{2}{c|}{\textbf{Bodmas}} & \multicolumn{2}{c|}{\textbf{Androzoo}} \\
\cline{2-5}
& \textbf{Time Range} & \textbf{\# of Malware : Benignware} & \textbf{Time Range} & \textbf{\# of Malware : Benignware} \\
\hline
\textbf{Initial Training Set}    & 2007-01 to 2019-09 & 4741 : 17565 & 2019-01 to 2019-12 & 4542 : 40947 \\
\hline
\textbf{Future Selection Set}  & 2019-11 & 2494 : 3718 & 2020-02 & 406 : 3697 \\
\hline
\end{tabular} }
\vspace{-20pt}
\end{table*}

The data drift detector finds samples in the \textbf{\textit{future selection set}} that deviate from the data distribution learned by the malware classifier, as the malware classifier can make wrong predictions when samples change over time. These drifted samples filtered by the data drift detector can be inspected by human for correction. Since it is costly for human to analyze samples, the victim can only manually investigate at most \textbf{\textit{k samples}} from the \textbf{\textit{future selection set}} (See Table~\ref{tab:bodmas_androzoo_dataset}). We assume the victim manually examines the $k$ most drifted samples picked by the initial data drift detectors~\cite{chen2023continuous}, as these drifted samples are mostly likely wrongly predicted by the malware classifier. Any malware among the $k$ selected samples will be analyzed by a human and blocked. This means for a successful attack, a malware must be \textbf{\textit{BOTH predicted as benign by the malware classifier AND not among the top drifted samples selected by the data drift detector}}.

Our \textbf{\textit{future selection set}} contains all samples two months after initial training (Table~\ref{tab:bodmas_androzoo_dataset}). For the poisoning attack, we pick this month (2019-11/2020-02) instead of simply the next month after initial training (2019-10/2020-01), as we assume it needs time from the next month to retrain the models. For comparable evalution results across all our evasion and poisoning attack experiments, we use the same future selection set for the evasion attack.

\scbf{Retrain/Poison Victim Models}: As time passes, the victim needs to update the malware classifier and data drift detector to combat concept drift. The new training set to retrain the malware classifier and data drift detector is formed by inserting the most drifted samples encountered by the existing data drift detector into the balanced \textit{initial training set}~\cite{chen2023continuous}. To only study the effect of poisoning and rule out other factors, our experiments assume all samples inserted to form the new training set are poisoning samples from the attacker for experimental purposes. For simplicity, we assume no further data balancing is done on the inserted samples and these updated models are trained from scratch.

We assume the victim conducts a rough manual triage on every inserted sample and correctly assigns every true label~\cite{yang2023jigsaw}. An attacker cannot insert any sample into the training set with a wrong label, and thus has to perform \textbf{\textit{clean-label poisoning}}. However, we assume the attacker can insert poisoning samples derived from benignware with benign functionality and labels~\cite{severi2021explanation}, as the victim includes recent samples from third-party intelligence platforms for training. All users, including the attacker, can upload files to these platforms, and the victim cannot carefully examine every incoming training sample to notice subtle changes, due to the sheer volume of incoming data.

\scbf{Attacker Knowledge}: We assume the attacker does not have full access to the victim models. However, as information for well-known malware families and benign software is semipublic, the attacker knows 10\% of benignware and 10\% of malware from the \textit{initial training set} of the victim model (See Table~\ref{tab:bodmas_androzoo_dataset}). The attacker builds \textbf{\textit{surrogate models}} according to their limited knowledge, and then attack the surrogate models they have full control over, hoping these attacks can transfer to the victim model.

\scbf{Feature Restrictions to Keep Malware Functionality}: For a practical attack, the feature values of a modifed sample cannot affect its functionality or file format. For example, adding unused space at the end of a Portable Executable (PE) file does not change its functionality, but changes its features and thus predictions from machine learning models.

We assume the attacker can increase feature values but cannot decrease them, as it is easy to increase these features (make file larger, add extra API calls, etc.) but cannot decrease them without breaking file functionality. As the features are properties of real-world software, all features also must be non-negative. For Bodmas~\cite{yang2021bodmas}, only 17 out of 2,381 features are feasible for modification without affecting the functionality of the underlying binary file~\cite{severi2021explanation} or breaking the file format. We assume there is no such restriction for Androzoo~\cite{allix2016androzoo}, and all of its 16,978 features can be incremented to 1.

\scbf{Perturbation Limit}: To preserve malicious functionality and avoid being conspicuous, the attacker needs resources to tweak files, so their corresponding features are changed. With limited resources, the universal perturbation added to malware/benignware cannot be arbitrarily large.

With Bodmas~\cite{yang2021bodmas}, different features have different value ranges. We scale all its features in the initial training data to be between 0 and 1 with a min-max scaler. We then set limits on the magnitude of the perturbation after scaling for its 17 modifiable features~\cite{severi2021explanation}. After the perturbation is added, if the magnitude of a feature exceeds the 0-to-1 limit, then this feature is clipped to 0 or 1.

With Androzoo~\cite{allix2016androzoo}, all features are either 0 or 1. As there are a large number of features (16,978 features) and the feature space is sparse, for better attack performance, we find significant features in Androzoo using Recursive Feature Elimination (RFE) with logistic regression~\cite{severi2021explanation}, while using data only known by the attacker rather tham the entire training set. We ensure only the selected significant features are modified. Depending on the number of modifiable features, we use the 200 and 1000 most significant features for a smaller and larger feature limit respectively.

\scbf{Attacker's Goals}: The attacker aims to attack both malware classification and data drift detection.

1) A malware sample crafted by the attacker should be classified as benignware by the malware classifier.

2) The malware sample should not be among the top-k drifting samples picked by the data drift detector so that it will not be inspected by the human analyst.

The attacker aims to compute a \textbf{\textit{universal perturbation}}, which can be generally applied to all malware samples to cause misclassification (See Section~\ref{subsec:attack_type}). Instead of computing one perturbation for every malware sample, this allows the attacker to compute the perturbation only once and apply to all.

\section{~\label{sec:attack_methods}Attack Methods}
\subsection{\label{subsec:attack_loss_func}Different Surrogates \& Loss for Perturbation}
While we study different methods to conduct evasion and poisoning attacks, the same sets of loss functions are used to generate perturbations in these attacks. We evaluate and compare perturbations generated from two methods. A perturbation can be generated from \textit{\textbf{either}} a surrogate MLP model \textit{\textbf{or}} a surrogate CAE model, using the respective loss function. In this approach, we do not combine the perturbations generated from the two different models.

\scbf{Option 1}: With a \textbf{\textit{surrogate MLP}} model, a perturbation $u$ can be generated to minimize the cross entropy loss of malware samples and the benign class:
\begin{equation}
\arg \min_{u} cross\_entropy(MLP(X_{malware} + u), y_{benign})
\end{equation}
While this perturbation is common for DNN classifiers, we also evaluate how it can evade data drift detection. We investigate whether moving malware samples across the decision boundary can simply make them a non-drifted member of the benign class.

\scbf{Option 2}: With a \textbf{\textit{surrogate CAE}} model, the crafted perturbation $u$ minimizes the distance between malware samples and benign centroid in the embedding space of the CAE:
\begin{equation}
    \begin{split}
    \arg \min_{u} MSE( CAE\_encoder(X_{malware} + u), benign\_centroid )
    \end{split}
\end{equation}
This is done because, in the latent space of the CAE model, a sample closer to the benign centroid is also more likely to be in the distribution of the benign class. We thus also evaluate how effective this perturbation is in moving malware samples across the decision boundaries of malware classifiers.

Algorithm\ref{alg:find_perturb} uses the above loss functions to compute the perturbation. It uses the rectified Adam optimizer and applies the \textit{perturbation limit} from Section~\ref{sec:general_threat_model_universal_attack}. To allow comparison, \textbf{\textit{the same perturbation}} is used for both evasion and poisoning.

\begin{algorithm}
\caption{\label{alg:find_perturb}Find Perturbation for Evasion and/or Poisoning}
\begin{algorithmic}[1]
\Procedure{find\_perturbation}{train\_dataset, loss\_type, model}
    \State u = $init\_zeros()$
    \For{round\_num in rounds}
        \For{$X_{malware}$ in load\_batch(train\_dataset)}
            \If{loss\_type == mlp}
                \State MLP = model
            	\State curr\_loss = $\min_{u} cross\_entropy(MLP(X_{malware} + u), y_{benign})$
            \Else
                \State CAE\_encoder, benign\_centroid = model \algorithmiccomment{The surrogate CAE model contains a centroid for its benign class, along with its encoder.}
                \State curr\_loss = $\min_{u} MSE(CAE\_encoder(X_{malware} + u), benign\_centroid )$
            \EndIf
            \State u = $RAdam.minimize\_backward(curr\_loss.mean())$
            \State u = $apply\_perturbation\_limit(u)$
        \EndFor
    \EndFor
    \State return u
\EndProcedure
\end{algorithmic}
\end{algorithm}

\subsection{\label{subsec:attack_type}Evasion and Poisoning Attacks}
Using the above loss functions, we conduct two different types of attacks: evasion and poisoning.

\scbf{Scenario 1) Evasion Attack}: The attacker adds a small perturbation to malware samples to cause misclassification. A perturbation can be generated for many samples or for each sample individually. A \textbf{\textit{universal perturbation}} aims to fool the model when added to any sample from a certain class and is computed using a pool of samples the attacker knows. This is different from a per-sample perturbation that aims to allow misclassification when added to only one specific sample but not other samples.

Our work focuses on \textbf{\textit{universal perturbation}}, as it is similar to the backdoor trigger in the sense that they are both generated from a large set of samples owned by the attacker and can be applied generally. While we also performed evaluation on per-sample perturbation, it has similar results as the universal perturbation.

\scbf{Scenario 2) Poisoning Attack}: The attacker injects poisoning samples into the training set so samples will be misclassified by the trained victim models during inference. We conduct \textbf{\textit{clean-label poisoning}} attacks, in which the attackers cannot change the correct labels for the injected training data to other labels they desire (restriction from Section~\ref{sec:general_threat_model_universal_attack}).

The attack is also \textbf{\textit{backdoor poisoning}}, during which the attacker generates a \textbf{\textit{backdoor trigger}} (same as the universal perturbation from Algorithm~\ref{alg:find_perturb} for evasion) and adds the same backdoor trigger to benign samples to form poisoning samples. The victim model trained with these poisoning samples should still have a generally good performance on clean inputs without the backdoor trigger. However, the poisoned victim model should misclassify a set of the attacker's samples only when the backdoor trigger is added to them. This increases stealthiness, as misclassification does not occur when no backdoor trigger is present in the sample.

The set of candidate poisoning samples is formed by adding the backdoor trigger to all benign samples known by the attacker. From Section~\ref{sec:general_threat_model_universal_attack}, as the victim inserts the most drifted samples it has seen to form the new training set, we assume only the most drifted poisoning samples from the candidate set will be inserted to retrain and poison the models.

\section{Evaluation}
We evaluate the attacks in Section~\ref{sec:attack_methods} with attacker capabilities, restrictions, and data described in Section~\ref{sec:general_threat_model_universal_attack}.

For generalizable attacks, when calculating the attack success rate (ASR) for testing, we apply the universal perturbation/backdoor trigger to \textbf{\textit{all malware samples}} in the \textit{initial training set} (Table~\ref{tab:bodmas_androzoo_dataset}) instead of only the malware samples the attacker knows. Experiments that apply universal perturbations to malware samples in the \textit{future selection set} have similar results and are not included in this paper. To eliminate randomness, we perform the attack for 30 runs and average the results across all runs.

We evaluate the impact of different model architecture on the victim and surrogate models. For MLP models, we use two hidden layers of 100 neurons between their input and output layers (denoted as 100-100). For CAE models, we evaluate a larger model structure of 512-384-256-128~\cite{chen2023continuous} and a smaller model structure of 512-128-32-8, where the layers in the encoder gradually decrease from 512 dimensions to 128-dimension and 8-dimension embedding space respectively. The layers in the CAE decoder is symmetric to the encoder with gradually increasing dimensions.

\subsection{Evaluation Metrics}
We compute an overall attack success rate $ASR_{overall}$ as our general evaluation metric. This overall attack success rate metric indicates how successful the attack is against both malware classification and data drift detection. The following shows how attack success against malware classification and data drift detection is defined to evaluate $ASR_{overall}$:

\textit{To attack against \textbf{malware classification}},  a crafted malware sample $m$ is successful if it is misclassified as benignware: $classification\_success(m) = (MLP\_prediction(m) == y_{benign})$. To evaluate the success of this subtask, we compute $ASR_{malware\_classification}$, which is simply the proportion of malware samples successfully misclassified as benignware.

\textit{To attack against \textbf{data drift detection}}, the crafted malware sample should not be among the top-k samples picked by the data drift detector. In most of our experiments, the number of top-drifting samples picked (k) is 500 for comparable results, and one experiment uses k = 200 to demonstrate patterns. A larger k is harder to evade, as more samples are selected and inspected by the human analyst.

Different data drift detectors use different metrics to compute the driftedness score and to pick drifting samples. As discussed in Section~\ref{sec:background_data_drift}, CADE picks top-drifting samples with the \textbf{\textit{highest}} OOD scores rather than the anomaly score of the benign class. For CADE, samples with higher OOD scores are considered more drifted. A crafted malware sample $m$ will evade CADE if its OOD score is lower than the kth most drifted sample in the \textbf{\textit{future selection set}}:

$drift\_success(m) = CADE\_OOD(m) < CADE\_OOD(s_{k})$, where
\begin{itemize}
    \item $s_{k} \in future\_selection\_dataset$
    \item $|\{CADE\_OOD(s) \ge CADE\_OOD(s_{k}) | s \in future\_selection\_dataset\}| = k$
\end{itemize}

Transcendent picks samples with the \textbf{\textit{lowest}} cred+conf score of the predicted class rather than the benign class.  A malware sample will evade Transcendent if its cred+conf score is higher than the cred+conf score of the kth most drifted sample in the \textbf{\textit{future selection set}}:
$drift\_success(m) = (cred+conf(m) > cred+conf(s_{k})$, where
\begin{itemize}
    \item $s_{k} \in future\_selection\_dataset$
    \item $|\{cred+conf(s) \le cred+conf(s_{k}) | s \in future\_selection\_dataset\}| = k$
\end{itemize}

\subsection{Evasion}
\scbf{Evading Malware Classification Only}: As a reference for our later experiments, Table~\ref{tab:universal_malevasion_mlp_clf} shows how our loss functions in Section~\ref{subsec:attack_loss_func} can evade a victim MLP model for malware classification only. For both Bodmas and Androzoo, the ASR against malware classification generally increases for perturbations from both surrogate CAE and surrogate MLP models, as the perturbation limit increases. This means our loss functions from both our surrogate models work against malware classification.

\begin{table}
\centering
\caption{\label{tab:universal_malevasion_mlp_clf} Evasion with Universal Perturbation Against MLP Malware Classification}
\begin{tabular}{|c|c|c|c|c|}
\hline
 & & \multicolumn{3}{c|}{ \textbf{ Mean $ASR_{malware\_classification}$ (Std) }} \\
\cline{3-5}
\textbf{Dataset} & \textbf{Perturbation Limit} & \multicolumn{3}{c|}{ \textbf{ Surrogate Model Structure }} \\
\cline{3-5}
 & & \textbf{MLP 100-100} & \textbf{CAE 512-384-256-128} & \textbf{CAE 512-128-32-8} \\
\hline
\multirow{4}*{Bodmas} & 0.25 & 77.7\% (3.2\%) & 79.1\% (5.1\%) & 78.5\% (6.3\%) \\
                      & 0.5  & 97.6\% (1.9\%) & 97.4\% (3.8\%) & 92.5\% (6.0\%) \\
                      & 0.75 & 100\% (0)      & 99.985\% (0.07\%) & 95.4\% (5.8\%) \\
                      & 1    & 100\% (0)      & 99.997\% (0.02\%) & 95.6\% (5.6\%) \\
\hline
\multirow{4}*{Androzoo} & 50 (Sig Feat 200)   & 72.1\% (2.8\%) & 71.2\% (2.3\%) & 72.2\% (2.7\%) \\
                        & 100 (Sig Feat 200)  & 74.3\% (2.9\%) & 71.6\% (2.5\%) & 73.6\% (2.7\%) \\
                        & 200 (Sig Feat 1000) & 98.3\% (0.7\%) & 95.0\% (1.8\%) & 95.3\% (2.0\%) \\
                        & 500 (Sig Feat 1000) & 98.9\% (0.4\%) & 94.9\% (1.9\%) & 95.4\% (1.9\%) \\
\hline
\end{tabular} 
\end{table}

\begin{figure}
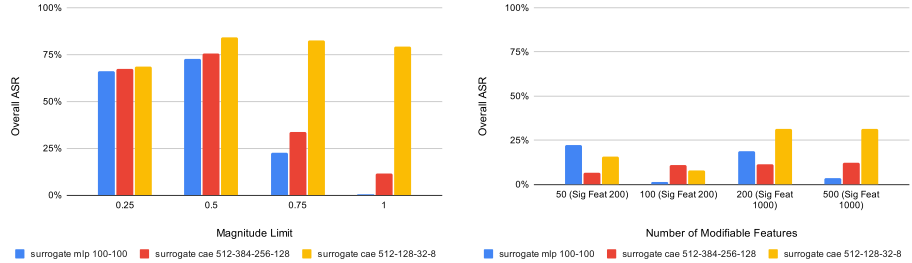

    \begin{subfigure}{0.5\textwidth}
         \includesvg[width=\linewidth]{figure_main/malevasion/cade_overall_asr_bodmas_tocae512-128-32-8_select500.svg}
         \caption{Bodmas, Top K = 500, Victim CAE 512-128-32-8}
         \label{fig:universal_malevasion_cade_bodmas_tocae512-128-32-8}
     \end{subfigure}
     \begin{subfigure}{0.5\textwidth}
         \includesvg[width=\linewidth]{figure_main/malevasion/cade_overall_asr_androzoo_tocae512-128-32-8_select500.svg}
         \caption{Androzoo, Top K = 500, Victim CAE 512-128-32-8}
         \label{fig:universal_malevasion_cade_androzoo_tocae512-128-32-8}
     \end{subfigure}
     \caption{\label{fig:universal_malevasion_cade} $ASR_{overall}$, Evasion Against Malware Classification + CADE Data Drift Detection}
     \vspace{-10pt}
\end{figure}

\scbf{Evading CADE + Malware Classification}: Unlike the malware classification case in Table~\ref{tab:universal_malevasion_mlp_clf}, when evading the CADE data drift detector and malware classifier in combination, the overall ASR can \textbf{\textit{decrease}} as an already large perturbation continues to increase, as shown in Figure~\ref{fig:universal_malevasion_cade}.

This is because the large perturbations result in large features in crafted samples. These large features form patterns different from other samples in the training set. The CAE models treat these large features as unseen patterns and push these samples away from the benign centroid. CADE with the underlying CAE models thus can mark these samples as drifting, resulting in a decreasing overall ASR. This means the attacker needs to modify features more carefully, as a feature capable of pushing samples towards the benign class in surrogate models can also have drifting side effects on the victim model.

This effect is best demonstrated from the Androzoo dataset in Figure~\ref{fig:universal_malevasion_cade}: with 200 most significant features selected by logistic regression (Section~\ref{sec:general_threat_model_universal_attack}), the overall ASR values for surrogate MLP and surrogate CAE 512-128-32-8 also decrease as the number of features increases from 50 to 100, as the selected features have more drifting side effect than the ability to push samples towards the benign class. However, after the number of significant features increases, with more features to select from, the attacker can pick more useful features, and the overall ASR values can increase again as the number of modifiable features increases from 100 (Sig Feat 200) to 200 (Sig Feat 1000).

As a large perturbation keeps increasing, the ASR from the \textbf{\textit{surrogate MLP}} decreases faster than the ASR from the \textbf{\textit{surrogate CAE}} models. This is because the cross-entropy loss for surrogate MLP does not take into account the distance towards the benign centroid in CAE, and therefore, perturbations generated from surrogate MLP prefer a region with higher confidence to be benign for MLP, but this region can be further away from the benign centroid of the CAE.

\begin{figure}[!htb]
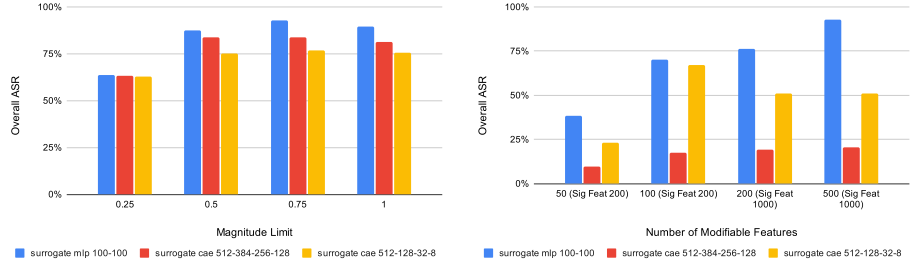
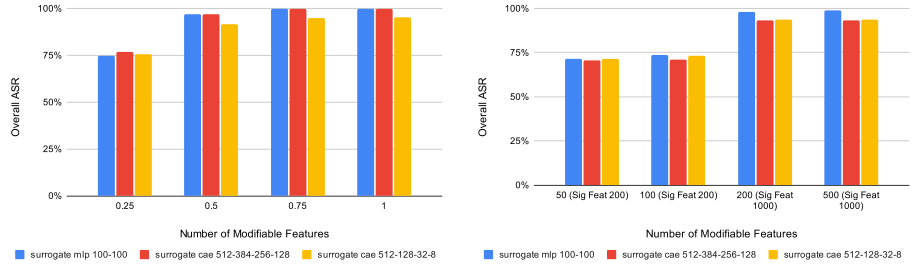

    \begin{subfigure}{0.5\textwidth}
         \includesvg[width=\linewidth]{figure_main/malevasion/transenc_overall_asr_bodmas_tocae512-128-32-8_select500.svg}
         \caption{Bodmas, Top K = 500, Victim CAE 512-128-32-8}
         \label{fig:universal_malevasion_transcend_bodmas_tocae512-128-32-8}
     \end{subfigure}
     \begin{subfigure}{0.5\textwidth}
         \includesvg[width=\linewidth]{figure_main/malevasion/transenc_overall_asr_androzoo_tocae512-128-32-8_select500.svg}
         \caption{Androzoo, Top K = 500, Victim CAE 512-128-32-8}
         \label{fig:universal_malevasion_transcend_androzoo_tocae512-128-32-8}
     \end{subfigure}
     \begin{subfigure}{0.5\textwidth}
         \includesvg[width=\linewidth]{figure_main/malevasion/transclf_overall_asr_bodmas_tomlp_select500.svg}
         \caption{Bodmas, Top K = 500, Victim MLP 100-100}
         \label{fig:universal_malevasion_transcend_bodmas_tomlp}
     \end{subfigure}
     \begin{subfigure}{0.5\textwidth}
         \includesvg[width=\linewidth]{figure_main/malevasion/transclf_overall_asr_androzoo_tomlp_select500.svg}
         \caption{Androzoo, Top K = 500, Victim MLP 100-100}
         \label{fig:universal_malevasion_transcendent_androzoo_mlp}
     \end{subfigure}
     \caption{\label{fig:universal_malevasion_transcendent} $ASR_{overall}$, Evasion Against Malware Classification + Transcendent Data Drift Detection}
     \vspace{-10pt}
\end{figure}

\scbf{Evading Transcendent + Malware Classification}: Unlike CADE with the CAE model, for Transcendent + CAE, large perturbations generated with the cross-entropy loss and the \textbf{\textit{surrogate MLP}} can achieve higher overall ASRs than the ones with \textbf{\textit{surrogate CAE}} models, as demonstrated in Figure~\ref{fig:universal_malevasion_transcend_bodmas_tocae512-128-32-8} and Figure~\ref{fig:universal_malevasion_transcend_androzoo_tocae512-128-32-8}. This is because Transcendent builds SVM on the CAE embedding space and uses distance to the decision boundary of the SVM as the metric to detect drifting samples. A drifting sample in CADE far away from the benign centroid can also have a large distance to the decision boundary and thus can be a non-drifting sample in Transcendent. As Transcendent takes the classification decision boundary into account, the perturbations from the \textbf{\textit{surrogate MLP}} have better ASRs in Transcendent than in CADE because these perturbations aim to push samples into regions with a high likelihood of being benign and, thus, more likely to be away from the decision boundary.

Also, Figure~\ref{fig:universal_malevasion_transcend_bodmas_tomlp} shows the overall ASR against Transcendent + MLP for Bodmas. Compared to Transcendent + CAE, Transcendent + MLP has much higher overall ASRs, and the universal perturbations generated from different CAE models also have smaller differences in $ASR_{overall}$. This is because, unlike the embedding space of victim CAE, the latent space of the victim MLP is not trained to deliberately enforce large distances between benignware and malware samples. Therefore, this makes it easier to move malware samples towards the benign class in Transcendent + MLP, and the advantages or disadvantages of universal perturbations from different surrogate models are not as significant.

\subsection{Clean-Label Backdoor Poisoning}
\begin{figure}[!htb]
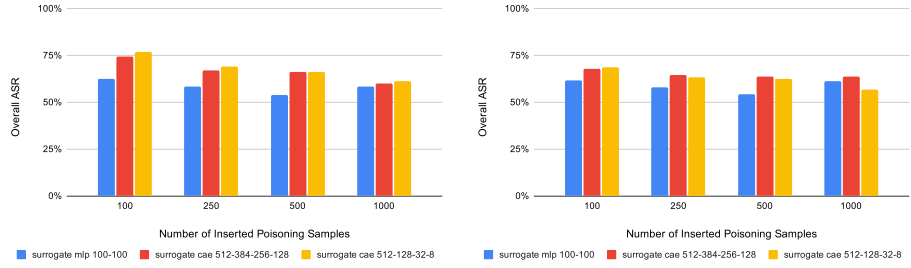

    \begin{subfigure}{0.5\textwidth}
         \includesvg[width=\linewidth]{figure_main/malpoison/malpoison_overall_asr_ood_bodmas_tocae512-128-32-8_poisonnum_clipnoise1_select500.svg}
         \caption{Top K = 500, Poisoned CADE 512-128-32-8}
         \label{fig:}
     \end{subfigure}
     \begin{subfigure}{0.5\textwidth}
         \includesvg[width=\linewidth]{figure_main/malpoison/malpoison_overall_asr_transenc_bodmas_tocae512-128-32-8_poisonnum_clipnoise1_select500.svg}
         \caption{Top K = 500, Poisoned Transcendent CAE 512-128-32-8}
         \label{fig:}
     \end{subfigure}
     \caption{\label{fig:universal_malpoison_bodmas} $ASR_{overall}$, Backdoor Poisoning Against Malware Classification and Data Drift Detection, Bodmas, Magnitude Limit = 1}
     \vspace{-10pt}
\end{figure}

\begin{figure}[!htb]
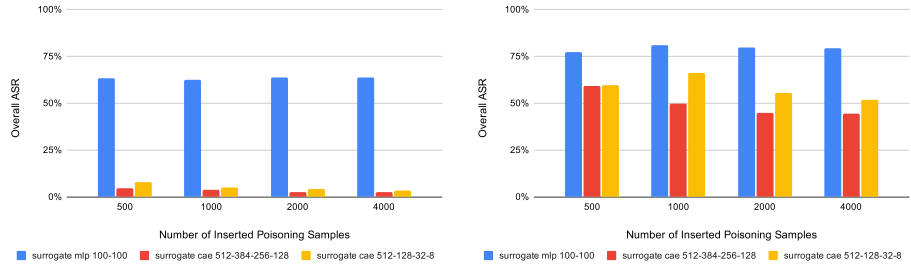

    \begin{subfigure}{0.5\textwidth}
         \includesvg[width=\linewidth]{figure_main/malpoison/malpoison_overall_asr_ood_androzoo_tocae512-128-32-8_poisonnum_feat500_select500.svg}
         \caption{\label{fig:universal_malpoison_androzoo_ood_cae512-128-32-8}Top K = 500, Poisoned CADE 512-128-32-8}
    \end{subfigure}
    \begin{subfigure}{0.5\textwidth}
         \includesvg[width=\linewidth]{figure_main/malpoison/malpoison_overall_asr_transenc_androzoo_tocae512-128-32-8_poisonnum_feat500_select200.svg}
         \caption{\label{fig:universal_malpoison_androzoo_transcendent_cae512-128-32-8_topk200}\textbf{\textit{Top K = 200}}, Poisoned Transcendent CAE 512-128-32-8}
    \end{subfigure}
    \caption{\label{fig:universal_malpoison_androzoo} $ASR_{overall}$, Backdoor Poisoning Against Malware Classification and Data Drift Detection, Androzoo, 500 Features (Sig Feat 1000)}
    \vspace{-10pt}
\end{figure}

Results from poisoning data drift detectors are also different than poisoning the malware classifier only (See Appendix). For both CADE and Transcendent + CAE, Figure~\ref{fig:universal_malpoison_bodmas} and \ref{fig:universal_malpoison_androzoo} show there can be a \textbf{\textit{decrease}} in the overall ASR with more poisoning samples. Note Figure~\ref{fig:universal_malpoison_androzoo_transcendent_cae512-128-32-8_topk200} uses a lower victim budget of \textbf{\textit{k = 200}} to better demonstrate patterns across the different number of inserted samples (See Appendix for same evaluation with k = 500).

This is due to the reconstruction loss term of the victim CAE models. As more poisoning samples are inserted, the CAE models need to differentiate between the different poisoning samples in the embedding space in order to reconstruct them accurately. However, the space in the embedding space close to the benign centroid is limited. As more benign poisoning samples are added, there is not enough space in the embedding space to reconstruct/differentiate different poisoning samples properly. The victim CAE models thus learn to put samples with the backdoor trigger into areas further away from the benign centroid (See Appendix). This causes the crafted samples to move away from the benign centroid, resulting in a decreasing overall ASR.

This effect is especially obvious when evaluated with the victim CAE model 512-128-32-8 that has a smaller embedding space and/or with the Androzoo dataset with more samples and input features, as the small capacity of the embedding space is less likely to be enough for the rich information in the input space. In such case as illustrated by Figure~\ref{fig:universal_malpoison_androzoo_ood_cae512-128-32-8}, poisoning with \textbf{\textit{surrogate CAEs}} has much lower ASRs than \textbf{\textit{surrogate MLPs}}, as the loss terms for CAE deliberately move poisoning samples towards the already crowded space around the benign centroid.

\section{Related Work}
\scbf{Evasion and Poisoning Attacks}: Existing works~\cite{boutsikas2021evading, song2003automatic} modify malware samples to evade malware classifiers that take manually crafted features extracted by static analysis. There are also works~\cite{lucas2021malware, demetrio2021adversarial, kreuk2018deceiving, kreuk2018adversarial} that modify malware samples to evade classifiers that take raw binaries as input. In addition, Severi et al.~\cite{severi2021explanation} and Jigsaw Puzzle~\cite{yang2023jigsaw} poison malware classifiers with static features to misclassify malware samples that have injected backdoor triggers.

Our work also conducts evasion and poisoning attacks against models that take input features from static analysis. However, to our knowledge, no existing work has attacked malware data drift detectors. This is different from our attack model, where we attack both data drift detection and malware classification.

\scbf{Surrogate Models and Attack Transferability}: Existing works~\cite{yang2023jigsaw, severi2021explanation, suciu2019exploring} assume the attacker does not have full knowledge of the victim model and thus trains surrogate models. The surrogate models in these existing works have the same features as the victim model but usually a smaller training set and/or different model architecture. Surrogate models in our work are built in similar ways.

\scbf{Keeping Malware Functionality}: Existing works modify malware samples to bypass machine learning models that use features extracted from cheap static analysis without the need to execute malware. The modified malware samples must keep the malicious functionalities and the file formats must not be broken.

Severi et al.\cite{severi2021explanation} and RAMEn~\cite{demetrio2021adversarial} perform operations on malware samples to change manually crafted static features that do not affect file format or functionality. Jigsaw Puzzle~\cite{yang2023jigsaw} adds bytecode gadgets from benign apps to Android malware while keeping the malicious behaviours intact. Lucas et al.~\cite{lucas2021malware} manipulates instructions that are part of the malware binary in a way that does not affect the execution result. Our work also modifies malware without affecting its functionality.

\scbf{Effective Attacks}: Existing works~\cite{yang2023jigsaw, severi2021explanation, lucas2021malware, demetrio2021adversarial, kreuk2018adversarial} for both evasion and poisoning can achieve high attack success rates. This is similar to the results in our work with high attack success rates.

To investigate the underlying reason for such success, Demetrio et al.~\cite{demetrio2019explaining} find that malware classifiers taking raw binary files as inputs learn relations between the class label and bytes that do not indicate true maliciousness. Suciu et al.~\cite{suciu2018does} further show that knowledge of the features used by the victim model is important for the success of evasion and poisoning attacks.

\section{Conclusion}
In conclusion, data drift detectors with underlying CAE models can be tricky to attack, as their embedding space and loss terms have complex interactions with attack parameters. Although larger perturbations for evasion are generally more effective against MLP malware classifiers, when evading CADE data drift detectors, a larger perturbation size pushes the crafted malware samples away from the benign centroid of CAE, dropping the attack success rate. More poisoning samples can inhibit attack success against data drift detectors with CAE, because the poisoned CAE can move samples with the backdoor trigger further away from the benign centroid, as it needs space to reconstruct the inserted poisoning samples. Additionally, the attacker's surrogate model and loss term need to be carefully chosen to align with the properties of the CAE embedding space under different scenarios.

\section{Acknowledgements}
Funding for this work was provided in part by NSERC Discovery Grant RGPIN-2026-07548 and NSERC-CSE Research Communities Grant ALLRP 588144-23. David Lie is supported by a Tier 1 Canada Research Chair CRC-2019-00242. Mingyue Yang has received support through Ontario Graduate Scholarships and Edward S. Rogers Sr. Graduate Scholarships. Researchers funded through the NSERC-CSE Research Communities Grants do not represent the Communications Security Establishment Canada or the Government of Canada. Any research, opinions or positions they produce as part of this initiative do not represent the official views of the Government of Canada.

\bibliographystyle{splncs04}
\bibliography{ref}

@inproceedings{yang2021cade,
  title={$\{$CADE$\}$: Detecting and explaining concept drift samples for security applications},
  author={Yang, Limin and Guo, Wenbo and Hao, Qingying and Ciptadi, Arridhana and Ahmadzadeh, Ali and Xing, Xinyu and Wang, Gang},
  booktitle={30th USENIX Security Symposium (USENIX Security 21)},
  pages={2327--2344},
  year={2021}
}

@inproceedings{jordaney2017transcend,
  title={Transcend: Detecting concept drift in malware classification models},
  author={Jordaney, Roberto and Sharad, Kumar and Dash, Santanu K and Wang, Zhi and Papini, Davide and Nouretdinov, Ilia and Cavallaro, Lorenzo},
  booktitle={26th USENIX security symposium (USENIX security 17)},
  pages={625--642},
  year={2017}
}

@inproceedings{barbero2022transcending,
  title={Transcending transcend: Revisiting malware classification in the presence of concept drift},
  author={Barbero, Federico and Pendlebury, Feargus and Pierazzi, Fabio and Cavallaro, Lorenzo},
  booktitle={2022 IEEE Symposium on Security and Privacy (SP)},
  pages={805--823},
  year={2022},
  organization={IEEE}
}

@inproceedings{allix2016androzoo,
  title={Androzoo: Collecting millions of android apps for the research community},
  author={Allix, Kevin and Bissyand{\'e}, Tegawend{\'e} F and Klein, Jacques and Le Traon, Yves},
  booktitle={Proceedings of the 13th international conference on mining software repositories},
  pages={468--471},
  year={2016}
}

@inproceedings{yang2021bodmas,
  title={BODMAS: An open dataset for learning based temporal analysis of PE malware},
  author={Yang, Limin and Ciptadi, Arridhana and Laziuk, Ihar and Ahmadzadeh, Ali and Wang, Gang},
  booktitle={2021 IEEE Security and Privacy Workshops (SPW)},
  pages={78--84},
  year={2021},
  organization={IEEE}
}

@article{anderson2018ember,
  title={Ember: an open dataset for training static pe malware machine learning models},
  author={Anderson, Hyrum S and Roth, Phil},
  journal={arXiv preprint arXiv:1804.04637},
  year={2018}
}

@inproceedings{severi2021explanation,
  title={$\{$Explanation-Guided$\}$ backdoor poisoning attacks against malware classifiers},
  author={Severi, Giorgio and Meyer, Jim and Coull, Scott and Oprea, Alina},
  booktitle={30th USENIX security symposium (USENIX security 21)},
  pages={1487--1504},
  year={2021}
}

@inproceedings{yang2023jigsaw,
  title={Jigsaw puzzle: Selective backdoor attack to subvert malware classifiers},
  author={Yang, Limin and Chen, Zhi and Cortellazzi, Jacopo and Pendlebury, Feargus and Tu, Kevin and Pierazzi, Fabio and Cavallaro, Lorenzo and Wang, Gang},
  booktitle={2023 IEEE Symposium on Security and Privacy (SP)},
  pages={719--736},
  year={2023},
  organization={IEEE}
}

@inproceedings{chen2023continuous,
  title={Continuous learning for android malware detection},
  author={Chen, Yizheng and Ding, Zhoujie and Wagner, David},
  booktitle={32nd USENIX Security Symposium (USENIX Security 23)},
  pages={1127--1144},
  year={2023}
}

@article{boutsikas2021evading,
  title={Evading malware classifiers via monte carlo mutant feature discovery},
  author={Boutsikas, John and Eren, Maksim E and Varga, Charles and Raff, Edward and Matuszek, Cynthia and Nicholas, Charles},
  journal={arXiv preprint arXiv:2106.07860},
  year={2021}
}

@article{song2003automatic,
  title={Automatic Generation of Adversarial Examples for Interpreting Malware Classifiers},
  author={Song, W and Li, X and Afroz, S and Garg, D and Kuznetsov, D and Yin, H},
  journal={arXiv preprint arXiv:2003.03100},
  year={2020}
}

@inproceedings{lucas2021malware,
  title={Malware makeover: Breaking ml-based static analysis by modifying executable bytes},
  author={Lucas, Keane and Sharif, Mahmood and Bauer, Lujo and Reiter, Michael K and Shintre, Saurabh},
  booktitle={Proceedings of the 2021 ACM Asia Conference on Computer and Communications Security},
  pages={744--758},
  year={2021}
}

@inproceedings{demetrio2021adversarial,
  title={Adversarial EXEmples: Functionality-preserving Optimization of Adversarial Windows Malware},
  author={Demetrio, Luca and Biggio, Battista and Lagorio, Giovanni and Armando, Alessandro and Roli, Fabio},
  booktitle={ICML 2021 Workshop on Adversarial Machine Learning},
  year={2021}
}

@article{kreuk2018deceiving,
  title={Deceiving end-to-end deep learning malware detectors using adversarial examples},
  author={Kreuk, Felix and Barak, Assi and Aviv-Reuven, Shir and Baruch, Moran and Pinkas, Benny and Keshet, Joseph},
  journal={arXiv preprint arXiv:1802.04528},
  year={2018}
}

@article{kreuk2018adversarial,
  title={Adversarial examples on discrete sequences for beating whole-binary malware detection},
  author={Kreuk, Felix and Barak, Assi and Aviv-Reuven, Shir and Baruch, Moran and Pinkas, Benny and Keshet, Joseph},
  journal={arXiv preprint arXiv:1802.04528},
  pages={490--510},
  year={2018}
}

@inproceedings{suciu2019exploring,
  title={Exploring adversarial examples in malware detection},
  author={Suciu, Octavian and Coull, Scott E and Johns, Jeffrey},
  booktitle={2019 IEEE Security and Privacy Workshops (SPW)},
  pages={8--14},
  year={2019},
  organization={IEEE}
}

@article{demetrio2019explaining,
  title={Explaining vulnerabilities of deep learning to adversarial malware binaries},
  author={Demetrio, Luca and Biggio, Battista and Lagorio, Giovanni and Roli, Fabio and Armando, Alessandro},
  journal={arXiv preprint arXiv:1901.03583},
  year={2019}
}

@inproceedings{suciu2018does,
  title={When does machine learning $\{$FAIL$\}$? generalized transferability for evasion and poisoning attacks},
  author={Suciu, Octavian and Marginean, Radu and Kaya, Yigitcan and Daume III, Hal and Dumitras, Tudor},
  booktitle={27th USENIX Security Symposium (USENIX Security 18)},
  pages={1299--1316},
  year={2018}
}

@inproceedings{arp2014drebin,
  title={Drebin: Effective and explainable detection of android malware in your pocket.},
  author={Arp, Daniel and Spreitzenbarth, Michael and Hubner, Malte and Gascon, Hugo and Rieck, Konrad and Siemens, CERT},
  booktitle={Ndss},
  volume={14},
  pages={23--26},
  year={2014}
}

@inproceedings{raff2018malware,
  title={Malware detection by eating a whole exe},
  author={Raff, Edward and Barker, Jon and Sylvester, Jared and Brandon, Robert and Catanzaro, Bryan and Nicholas, Charles K},
  booktitle={Workshops at the thirty-second AAAI conference on artificial intelligence},
  year={2018}
}

\appendix

\section{\label{appen_sec:universal_evasion} Appendix} 

\begin{figure}[!htb]
    \begin{subfigure}{0.5\textwidth}
         \includesvg[width=\linewidth]{figure_main/malpoison/malpoison_overall_asr_ood_bodmas_tocae512-384-256-128_poisonnum_clipnoise1_select500.svg}
         \caption{Top K = 500, Poisoned CADE 512-384-256-128}
         \label{fig:}
     \end{subfigure}
     \begin{subfigure}{0.5\textwidth}
         \includesvg[width=\linewidth]{figure_main/malpoison/malpoison_overall_asr_transenc_bodmas_tocae512-384-256-128_poisonnum_clipnoise1_select500.svg}
         \caption{Top K = 500, Poisoned Transcendent 512-384-256-128}
         \label{fig:}
     \end{subfigure}
     \caption{\label{fig:universal_malpoison_tocae512-384-256-128_bodmas} $ASR_{overall}$, Backdoor Poisoning Against Malware Classification and Data Drift Detection, Bodmas, Magnitude Limit = 1, Poisoned CAE 512-384-256-128}
     \vspace{-20pt}
\end{figure}

\begin{figure}[!htb]
    \begin{subfigure}{0.5\textwidth}
         \includesvg[width=\linewidth]{figure_main/malpoison/malpoison_overall_asr_ood_androzoo_tocae512-384-256-128_poisonnum_feat500_select500.svg}
         \caption{Top K = 500, Poisoned CADE 512-384-256-128}
    \end{subfigure}
    \begin{subfigure}{0.5\textwidth}
         \includesvg[width=\linewidth]{figure_main/malpoison/malpoison_overall_asr_transenc_androzoo_tocae512-128-32-8_poisonnum_feat500_select500.svg}
         \caption{Top K = 500, Poisoned Transcendent CAE 512-128-32-8}
    \end{subfigure}
    \begin{subfigure}{0.5\textwidth}
         \includesvg[width=\linewidth]{figure_main/malpoison/malpoison_overall_asr_transenc_androzoo_tocae512-384-256-128_poisonnum_feat500_select500.svg}
         \caption{Top K = 500, Poisoned Transcendent CAE 512-384-256-128}
    \end{subfigure}
    \begin{subfigure}{0.5\textwidth}
         \includesvg[width=\linewidth]{figure_main/malpoison/malpoison_overall_asr_transenc_androzoo_tocae512-384-256-128_poisonnum_feat500_select200.svg}
         \caption{\textbf{\textit{Top K = 200}}, Poisoned Transcendent CAE 512-384-256-128}
    \end{subfigure}
    \caption{\label{appen_fig:universal_malpoison_androzoo_appendix} $ASR_{overall}$, Backdoor Poisoning Against Malware Classification and Data Drift Detection, Androzoo, 500 Features (Sig Feat 1000)}
    \vspace{-20pt}
\end{figure}

\begin{figure}
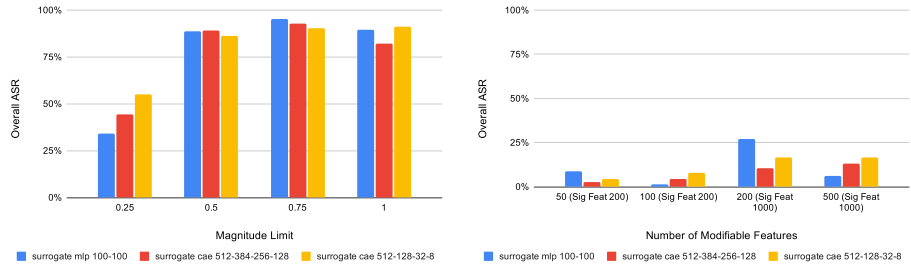
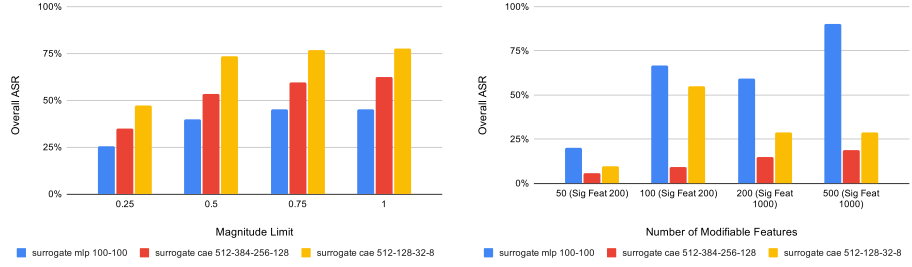

    \begin{subfigure}{0.5\textwidth}
         \includesvg[width=\linewidth]{figure_main/malevasion/cade_overall_asr_bodmas_tocae512-384-256-128_select500.svg}
         \caption{Bodmas, Top K = 500, Victim CADE 512-384-256-128}
         \label{fig:universal_malevasion_cade_bodmas_tocae512-384-256-128}
     \end{subfigure}
     \begin{subfigure}{0.5\textwidth}
         \includesvg[width=\linewidth]{figure_main/malevasion/cade_overall_asr_androzoo_tocae512-384-256-128_select500.svg}
         \caption{Androzoo, Top K = 500, Victim CADE 512-384-256-128}
         \label{fig:universal_malevasion_cade_androzoo_tocae512-384-256-128}
     \end{subfigure}
    \begin{subfigure}{0.5\textwidth}
         \includesvg[width=\linewidth]{figure_main/malevasion/transenc_overall_asr_bodmas_tocae512-384-256-128_select500.svg}
         \caption{Bodmas, Top K = 500, Victim Transcendent CAE 512-384-256-128}
         \label{fig:universal_malevasion_transcend_bodmas_tocae512-384-256-128}
     \end{subfigure}
     \begin{subfigure}{0.5\textwidth}
         \includesvg[width=\linewidth]{figure_main/malevasion/transenc_overall_asr_androzoo_tocae512-384-256-128_select500.svg}
         \caption{Androzoo, Top K = 500, Victim Transcendent CAE 512-384-256-128}
         \label{fig:universal_malevasion_transcend_androzoo_tocae512-384-256-128}
     \end{subfigure}
     \caption{\label{appen_fig:universal_malevasion_tocae512-384-256-128_androzoo} $ASR_{overall}$, Evasion Against Malware Classification and Data Drift Detection, Victim CAE 512-384-256-128}
     \vspace{-20pt}
\end{figure}

\begin{table*}
\centering
\caption{\label{tab:universal_malpoison_mlp_clf_noisesize} Varying Perturbation Limit for Backdoor Poisoning Against MLP Malware Classification Only}
\begin{tabular}{|c|c|c|c|c|}
\hline
\textbf{Dataset \&} & \textbf{Perturbation} & \multicolumn{3}{c|}{ \textbf{ Mean $ASR_{malware\_classification}$ }} \\
\cline{3-5}
\textbf{\# of Inserted} & \textbf{Limit} & \multicolumn{3}{c|}{ \textbf{ Surrogate Model Structure }} \\
\cline{3-5}
\textbf{Poisoning Samples} & & \textbf{MLP 100-100} & \textbf{CAE 512-384-256-128} & \textbf{CAE 512-128-32-8} \\
\hline
Bodmas                & 0.25  & 85.5\% & 82.1\% & 73.7\% \\
1000 Inserted         & 0.5   & 99.9\% & 94.9\% & 84.7\% \\
Poisoning Samples     & 0.75  & 100\%  & 100\%  & 90.1\% \\
                      & 1     & 100\%  & 100\%  & 91.3\% \\
\hline
Androzoo              & 50 (Sig Feat 200)   & 87.84\% & 86.92\% & 87.26\% \\
4000 Inserted         & 200 (Sig Feat 1000) & 99.92\% & 99.74\% & 99.79\% \\
Poisoning Samples     & 500 (Sig Feat 1000) & 99.98\% & 99.62\% & 99.65\% \\
\hline
\end{tabular}
\end{table*}

\begin{table*}
\centering
\caption{\label{tab:universal_malpoison_mlp_clf_poisonnum} Varying Number of Poisoning Samples for Backdoor Poisoning Against MLP Malware Classification Only}
\begin{tabular}{|c|c|c|c|c|}
\hline
\textbf{Dataset \&} & \textbf{\# of Inserted} & \multicolumn{3}{c|}{ \textbf{ Mean $ASR_{malware\_classification}$ }} \\
\cline{3-5}
\textbf{Perturbation} & \textbf{Poisoning Samples} & \multicolumn{3}{c|}{ \textbf{ Surrogate Model Structure }} \\
\cline{3-5}
\textbf{Limit} & & \textbf{MLP 100-100} & \textbf{CAE 512-384-256-128} & \textbf{CAE 512-128-32-8} \\
\hline
Bodmas                & 100  & 100\% & 99.7\% & 87.1\% \\
Magnitude Limit 1     & 200  & 100\% & 99.9\% & 88.2\% \\
                      & 500  & 100\% & 100\% & 90.0\% \\
                      & 1000 & 100\% & 100\% & 91.3\% \\
\hline
Androzoo              & 500  & 99.98\% & 99.34\% & 99.61\% \\
500 Features          & 1000 & 99.99\% & 99.51\% & 99.58\% \\
(Sig Feat 1000)       & 2000 & 99.98\% & 99.65\% & 99.70\% \\
                      & 4000 & 99.98\% & 99.62\% & 99.65\% \\
\hline
\end{tabular}
\end{table*}

\begin{table*}
\centering
\caption{\label{tab:universal_malpoison_cade_dist_bodmas_poisonnum} CAE Distance to Benign Centroid and Benign Anomaly Score for Backdoor Poisoning, Varying Number of Poisoning Samples, Bodmas, Magnitude Limit 1}
\begin{tabular}{|c|c|c|c|c|}
\hline
& & \multicolumn{3}{c|}{\textbf{Avg Dist to Benign Centroid / Avg Benign Anomaly Score}} \\
\cline{3-5}
\textbf{Poisoned CAE} & \textbf{\# Inserted} & \multicolumn{3}{c|}{\textbf{Surrogate Model Structure}} \\
\cline{3-5}
& \textbf{Poisoning Samples} & \textbf{MLP 100-100} & \textbf{CAE 512-384-256-128} & \textbf{CAE 512-128-32-8} \\
\hline
512-128-32-8    & 100  & 1.77 / 705 & 1.05 / 386 & 1.15 / 504 \\
                & 250  & 1.82 / 713 & 1.44 / 593 & 1.68 / 662 \\
                & 500  & 1.77 / 756 & 1.35 / 551 & 1.90 / 707 \\
                & 1000 & 1.67 / 724 & 1.78 / 635 & 2.33 / 1016 \\
\hline
512-384-256-128 & 100  & 0.172 / 2.49 & 0.229 / 3.66 & 0.631 / 15.2 \\
                & 250  & 0.194 / 2.80 & 0.232 / 4.16 & 0.718 / 16.2 \\
                & 500  & 0.163 / 2.18 & 0.170 / 2.46 & 0.703 / 15.8 \\
                & 1000 & 0.143 / 1.85 & 0.148 / 1.88 & 0.592 / 14.6 \\
\hline
\end{tabular}
\end{table*}

\begin{table*}
\centering
\caption{\label{tab:universal_malpoison_cade_dist_androzoo_poisonnum} CAE Distance to Benign Centroid and Benign Anomaly Score for Backdoor Poisoning, Varying Number of Poisoning Samples, Androzoo, 500 Features (Sig Feat 1000)}
\begin{tabular}{|c|c|c|c|c|}
\hline
& & \multicolumn{3}{c|}{ \textbf{Avg Dist to Benign Centroid / Avg Benign Anomaly Score} } \\
\cline{3-5}
\textbf{Poisoned CAE} & \textbf{\# Inserted} & \multicolumn{3}{c|}{\textbf{Surrogate CAE Structure}} \\
\cline{3-5}
& \textbf{Poisoning Samples} & \textbf{MLP 100-100} & \textbf{CAE 512-384-256-128} & \textbf{CAE 512-128-32-8} \\
\hline
512-128-32-8    & 500   & 1.03 / 31 & 1.97 / 63 & 1.88 / 61 \\
                & 1000  & 0.99 / 31 & 1.97 / 66 & 1.95 / 63 \\
                & 2000  & 0.89 / 28 & 2.12 / 73 & 1.92 / 62 \\
                & 4000  & 0.87 / 27 & 2.18 / 78 & 1.98 / 71 \\
\hline
512-384-256-128 & 500   & 0.79 / 25 & 1.72 / 59 & 1.72 / 57 \\
                & 1000  & 0.90 / 30 & 2.09 / 69 & 2.06 / 72 \\
                & 2000  & 0.90 / 32 & 2.29 / 81 & 2.30 / 82 \\
                & 4000  & 0.89 / 32 & 2.35 / 91 & 2.39 / 92 \\
\hline
\end{tabular}
\end{table*}

\end{document}